\documentstyle[twocolumn,prl,aps,psfig]{revtex}

\begin{document}
\draft
\preprint{USITP-98-19}

\title{Effective Action Studies of Quantum Hall Spin Textures}

\author{K. Lejnell$^1$, A. Karlhede$^{1}$ and S. L. Sondhi$^{2}$.}

\address{$^{1}$ Department of Physics, Stockholm University,
Box 6730, S-11385 Stockholm, Sweden}
\address{$^{2}$ Department of Physics, Princeton University,
Princeton NJ 08544}

\date{\today}
\maketitle

\begin{abstract}
We report on analytic and numerical studies of spin textures in quantum
Hall systems using a long-wavelength effective action for the magnetic
degrees of freedom derived previously. The majority of our results concern
skyrmions or solitons of this action. We have constructed approximate analytic
solutions for skyrmions of arbitrary topological and electric charge and
derived expressions for their energies and charge and spin radii. We 
describe a combined shooting/relaxational technique for numerical determination
of the skyrmion profiles and present results that compare favorably with
the analytic treatment as well as with Hartree-Fock studies of these objects.
In addition, we describe a treatment of textures at the edges of quantum Hall
systems within this approach and provide details not reported previously.

\end{abstract}


%\pacs{74.60Ec;74.75.+t????}

\section{Introduction}
\label{sec_intro}
Following the demonstration in \cite{skkr} that quantum Hall ferromagnets
contain skyrmions in their quasiparticle spectra, there has been a 
great deal of work on spin textures in quantum Hall systems. This 
work has received considerable impetus from a set of experiments
\cite{barrett,schmeller,goldberg,eaves} that have adduced 
evidence for skyrmions being the
lowest energy quasiparticles at $\nu=1$ as well as for their {\em not}
being so at $\nu=3$ and higher odd integer fillings \cite{wu,jain}.
The skyrmions themselves have been investigated by several authors
\cite{fertig,skothers} and a skyrme crystal phase has been studied
\cite{crystal} with a recent experiment appearing to find a transition
into it at a finite temperature \cite{bayot}. 
In addition, textured 
quasiparticles have been discussed in the context of double layer 
systems \cite{indiana} and invoked in explaining a novel phase transition 
in the latter \cite{murphy}. More recently, it has been shown that 
ferromagnetic quantum Hall states can exhibit textured 
edges \cite{karlhede,others}.

The analysis in \cite{skkr} was based on a long wavelength effective
action for the spin dynamics, supplemented by a charge density-topological
density constraint, that allowed an analytic computation of the properties 
of the skyrmions in the limit of small Zeeman energies. No details of this
computation were given there and one of the purposes of this paper is to 
provide them. A second is to report numerical studies of the effective 
action, that have a region of validity beyond the analytic approximations,
and compare extremely well with those using the Hartree-Fock technique of 
Fertig {\it et. al.} \cite{fertig}. We also report here on the details of
effective action studies of the textured edges mentioned in \cite{karlhede}.
We note that the numerical studies have been described previously in a thesis
by one of us \cite{kennet} and are being reported here for ease of
access. In the interim, effective action  results on skyrmion properties have 
been reported by Abolfath {\it et. al.} using methods similar to ours\cite{abolfath}, 
and by Rao {\it et. al.} using a variational approach \cite{rao},
while Moon and Mullen have presented an improved action
that is accurate even for small skyrmion sizes \cite{moon}.

The paper is organized as follows. After introducing the effective action in Sec. 
\ref{sec_eff_act},
we outline the analytic
derivation of skyrmion properties at small Zeeman energies in Sec. \ref{sec analytic}. 
In Sec. \ref{sec_relaxation},
we discuss the technique utilized in their numerical computation. In Sec. \ref{sec_results}, 
we present the numerical results for charge 1 and charge 2 skyrmions at $\nu =1$, and 
compare to Hartree-Fock calculations and analytic expansions. 
We then discuss the asymptotic shape of the 
skyrmions (Sec. \ref{sec_asy}). This is followed by results for 
textured edges (Sec. \ref{sec_edg}) and a brief summary (Sec. \ref{sec_sum}).

\section{Effective Action and Skyrmions}
\label{sec_eff_act}
In the long wavelength limit, the spin degree of freedom of a quantum 
Hall (QH) ferromagnet is described by a unit vector ${\bf n}({\bf r})$. 
A distinguishing feature of QH ferromagnets is that the topological density
of the spin field, $q({\bf r})={\bf n}\cdot (\partial _x {\bf n} \times 
\partial _y {\bf n})/4\pi$, is proportional to the deviation 
of the charge density, $\rho({\bf r})$, from its background value, 
$\rho-\bar \rho = \nu_{{\scriptscriptstyle FM}}q$ where 
$\bar \rho=\nu_{{\scriptscriptstyle FM}}/2\pi \ell^2$.
($\nu_{{\scriptscriptstyle FM}}$  is the filling factor of the ferromagnetic 
component of the QH liquid ({\it e.g.} $\nu_{{\scriptscriptstyle FM}}=1$ at 
$\nu=3$), and $\ell = \sqrt{\hbar c/eB}$ 
is the magnetic length.) This enables one to 
formally integrate out the dynamics of the charge
and obtain an effective Lagrangian for the system in terms of the spin
field alone whose coefficients can be fixed using known long wavelength
quantities \cite{skkr}:
\begin{eqnarray}
\label{Leff}
{\cal L}_{eff} &=&\frac 1 2 {\bar \rho} {\bf {\cal A}}({\bf n}) 
\partial_t {\bf n}
- \frac{1}{2}\rho_s (\nabla {\bf n})^2
\nonumber \\
&+& \frac{1}{2} g \overline{\rho} \mu_{\scriptscriptstyle B} {\bf n} {\bf B}
-\nu _{{\scriptscriptstyle FM}}^2 \frac{e^2}{2 \epsilon} 
\int d^2r' \frac{q({\bf r}) q({\bf r}')} {|{\bf r}-{\bf r}'|} \ \ \ .
\end{eqnarray}
Here, ${\bf {\cal A}}$ is the vector potential of a unit magnetic monopole, 
$\rho_s$ is the spin stiffness \cite{rho_s} and 
$\epsilon$  the dielectric constant of the background semiconductor. 
Technically, ${\cal L}_{eff}$
describes an $O(3)$ $\sigma-$model.

As we will be interested in this paper in static configurations of the field, 
we will seek to minimize the energy functional
\begin{equation}
\label{Eeff}
E = -\int d^2r [{\cal L}_{eff}-
\frac 1 2 {\bar \rho} {\bf {\cal A}}({\bf n}) 
\partial_t {\bf n} ] \equiv E_{G} + E_{\cal{Z}} + E_{C} \ \ \ ,
\end{equation}
where the gradient ($E_{G}$), the Zeeman ($E_{\cal{Z}}$) and the Coulomb 
($E_{C}$) energies can be read off from Eq. (\ref{Leff}).
In the following we will measure all energies in units of 
$e^2/\epsilon \ell$ and lengths in units of $\ell$.

For finite energy configurations, the finiteness of the gradient term
requires that the field approach a common value at infinity regardless
of direction. In such cases, the plane can be compactified to a sphere
and ${\bf n}({\bf r})$ gives a map from $S_2$ (the compactified plane)
to $S_2$ (spin space or the target space of the $\sigma-$model).
Such maps are characterized by an integer topological charge $Z$, which can 
be expressed as $Z=\int d^2r q({\bf r})$, where  $q$ is the topological 
(Pontryagin) density introduced earlier. As noted there, the topological 
density plays a crucial role in the physics of QH ferromagnets as it 
is proportional
to the charge density of the underlying itinerant system. Consequently,
localized configurations of topological charge $Z$, which we shall term
charge $Z$ skyrmions, carry electric charge $\nu_{{\scriptscriptstyle FM}}Z|e|$. 

In the absence of the Zeeman and 
Coulomb terms, the energy functional (\ref{Eeff}) is
scale invariant and rotationally invariant but on account of the 
non-linearity implicit in the definition of ${\bf n}$, finding its
skyrmion solutions is a non-trivial problem, solved previously by Belavin and 
Polyakov (BP) \cite{bp,rajaraman}. The additional terms break both 
symmetries although they preserve rotations about the field axis which we 
shall take to be the $z$-axis. For the full action, the BP technique (based
on achieving a Bogomol'nyi bound \cite{bogomolnyi} for the action) breaks 
down with the non-locality of the Coulomb term making matters even worse. 
Consequently, it is necessary to attack the Euler-Lagrange equation for 
the energy functional directly.

We now derive this equation for the skyrmion configurations by making an 
ansatz that has topological charge $Z$ and depends on one unknown function. 
Calculating the energy of this configuration and minimizing with respect
to the unknown function gives an integro-differential equation.

\begin{figure}
\begin{center}
\leavevmode
\psfig{file=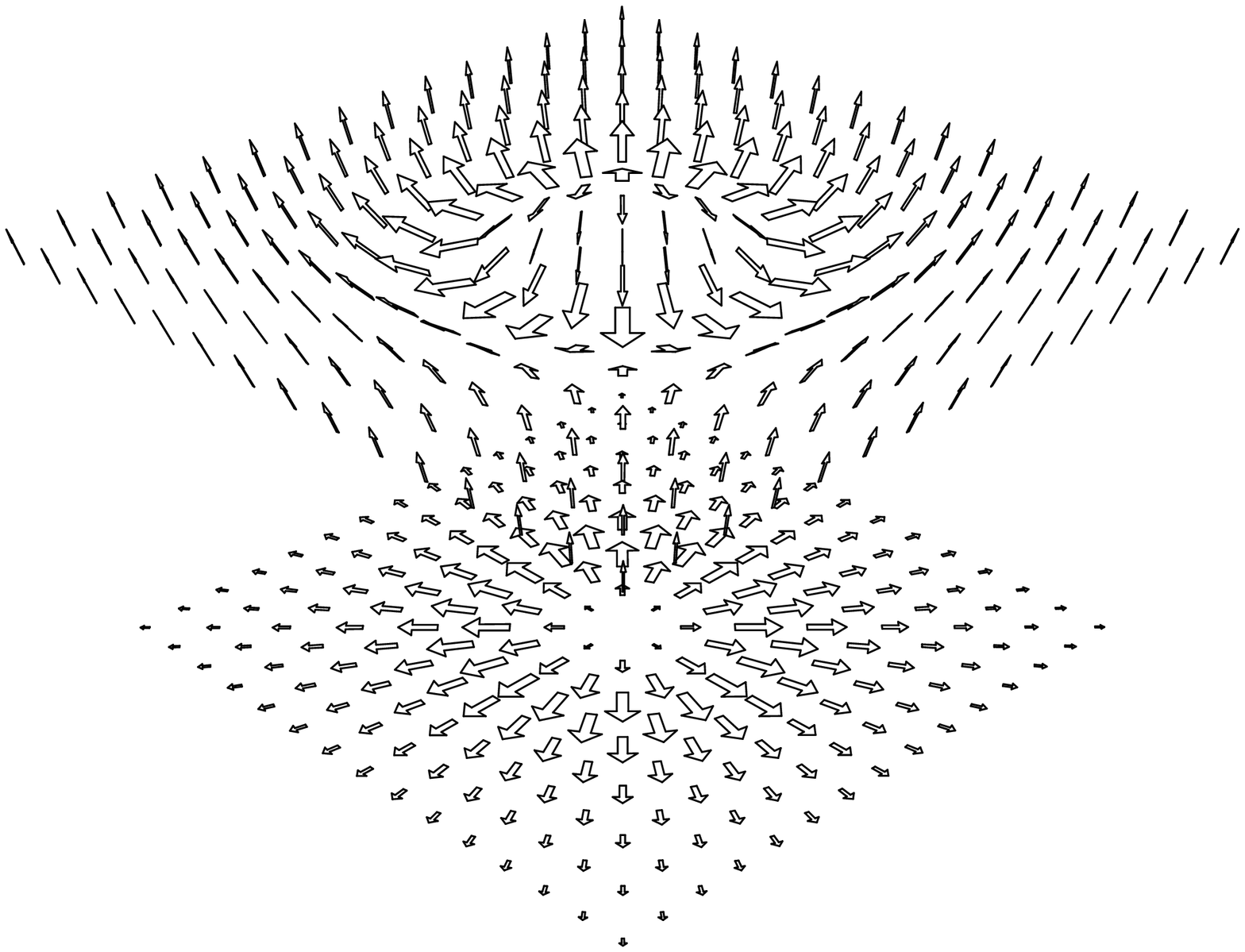,height=6.5cm,angle=-90}
\psfig{file=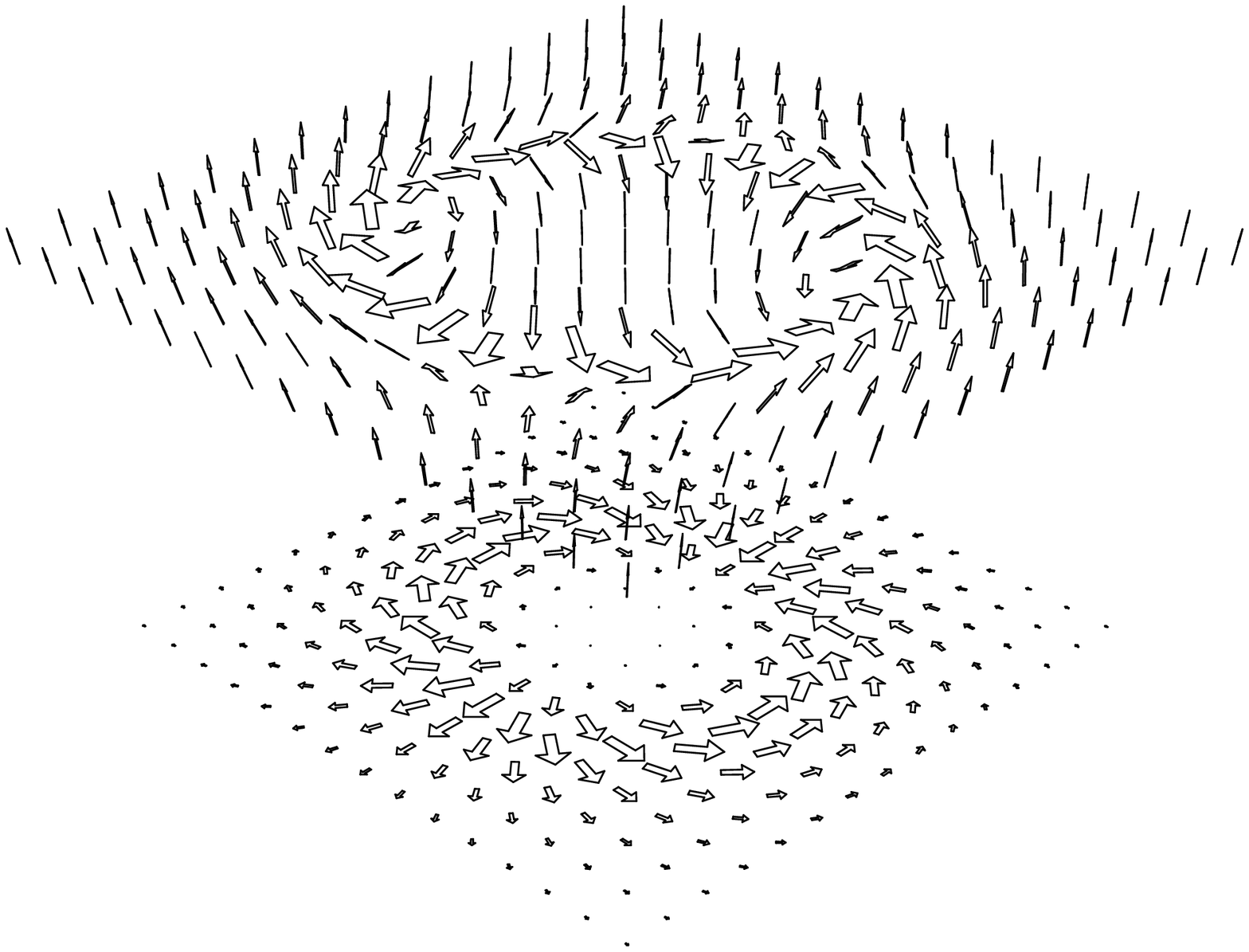,height=6.5cm,angle=-90}
\end{center}
\caption{Skyrmion spin vector field, $\bf{n}(\bf{r})$, 
and its projection onto the plane of the electron gas
for $Z=1$ (top figure) and $Z=2$ (bottom figure).}
\label{fig: xyspin}
\end{figure}

In the ground state the spin of the ferromagnetic component is polarized:
${\bf n}=\hat z$. For a skyrmion with topological charge $Z$, we make the 
ansatz (in polar coordinates  $(r,\theta)$ with their origin at the center
of the skyrmion): 
\begin{eqnarray}
\label{ansatz}
  n_x &=& \sqrt{1 - f^2(r)} \cos(Z\theta) \ \ , 
\nonumber \\
    n_y &=& \sqrt{1 - f^2(r)} \sin(Z\theta) \ \ , \nonumber \\
    n_z &=& f(r) \ \ . 
\end{eqnarray}
This leads to the topological density
\begin{equation}
\label{pontryagin}
q({\bf r})={Z \over 4 \pi r} {df \over dr} \ \ \ .
\end{equation}
For a finite energy configuration, the spin vector must
be aligned with the external magnetic field at infinity. At the 
center the spin points in the opposite direction. This leads to 
the boundary conditions  $f(0) = -1, \, f(\infty) = 1$, and (\ref{pontryagin})
shows  that (\ref{ansatz}) describes configurations with 
topological charge $Z$.  Note that the topological density is 
independent of $\theta$, {\it i.e.} it is spatially rotationally invariant; since
a shift $\theta \rightarrow \theta - \theta_o$ in (\ref{ansatz}) is a spin 
rotation 
it is clear that the topological density is invariant under spin rotations about 
the $z$-axis as well. Note however, that
the spin field itself is not separately invariant under rotations and
spin rotations --- it is invariant only under the combination generated
by $L_z + ZS_z$. It is easy to convince oneself that respecting both
symmetries is incompatible with non-trivial skyrmion solutions. 
Consequently, our ansatz is the maximally symmetric one possible. 
Examples of spin configurations for $Z=1$ and $Z=2$ are sketched in 
Fig. \ref{fig: xyspin}.

The function $f(r)$ in (\ref{ansatz}) is determined by 
minimizing the energy (\ref{Eeff}). 
We now calculate the energy for a skyrmion of the form (\ref{ansatz}). The 
gradient and Zeeman energies are:
\begin{eqnarray}
\label{sigma energy}
E_{G}&=&\pi \rho_s \int_0^{\infty}dr [Z^2\frac{1-f^2}{r} +
  \frac{rf'^2}{1-f^2}] \ \ \ ,
\\
\label{Zeeman energy}
E_{\cal{Z}} &=& -  \frac 1 2 \nu_{{\scriptscriptstyle FM}} \tilde{g} 
\int_0^{\infty}
dr r f(r) \ \ \ ,
\end{eqnarray}
where $\tilde g = g \mu_B B/(e^2/\epsilon \ell)$ and $f'=df/dr$.
We write the Coulomb energy as
\begin{equation}
  \label{Coulomb energy1}
  E_C= {1 \over 4} {\nu_{{\scriptscriptstyle FM}}^2 Z^2}\int_0^{\infty}dr 
f'V_C(r,f') \ \ \ ,
\end{equation}
where
\begin{eqnarray}
\label{Coulomb potential}
  V_C(r,f') =\frac 1 {4 \pi} \int_0^{\infty} dr' f'(r') 
\int_0^{2 \pi}d\theta'[r^2 + r'^2 \nonumber \\
-2 r r' \cos (\theta' - \theta)]^{-1/2} \nonumber \\
\nonumber \\
=\frac 1 { \pi} \int_0^{\infty} dr' \frac {f'(r')} {r+r'}  
\cdot F(\frac{\pi}{2},2 \frac{\sqrt{r r'}}{r+r'}) \ \ \ .
\end{eqnarray}
Here,  $F(\varphi,x)$ is the elliptic 
integral of the first kind \cite{convention}.

Adding all energy contributions, (\ref{sigma energy}), 
(\ref{Zeeman energy}) and (\ref{Coulomb energy1}), we obtain the 
final expression for the total energy:
\begin{eqnarray}
  \label{eq: total energy}
  E=\int_0^{\infty}dr[\pi \rho_s  (Z^2 \frac{1-f^2}{r} +
  \frac{rf'^2}{1-f^2})
\nonumber \\ 
  - \frac 1 2 \nu_{{\scriptscriptstyle FM}} \tilde{g} r
  f + {1\over4} \nu_{{\scriptscriptstyle FM}}^2 Z^2 f' V_C(r,f')] \ \ \ .
\end{eqnarray}
Varying the energy functional with respect to $f$ 
we find (observing that $\delta E_C/\delta f$ gives a contribution
$-2\partial V_C/\partial r$) the non-linear and non-local integro-differential 
equation:
\begin{eqnarray}
  \label{DE1}
  [r^2 f'' + r f'](1-f^2)
  +r^2 ff'^2
  + [Z^2 f+\frac{\tilde{g}\nu_{{\scriptscriptstyle FM}}} {4\pi \rho_s} r^2 
\nonumber \\
  + \frac { \nu_{{\scriptscriptstyle FM}}^2 Z^2 }{4 \pi \rho_s}
  r \frac{\partial
    V_C}{\partial r}] \cdot (1-f^2)^2 = 0 \ \ \ .
\end{eqnarray}
Note that $V_C$ is a functional of $f$.

As we noted earlier, the gradient energy, $E_{G}$, is scale invariant and 
that minimizing it alone leads to the BP solutions of arbitrary size 
$\lambda$. Rewriting Eq. (\ref{sigma energy}) as
\begin{equation}
E_{G}=\mp 4\pi \rho_s Z+ \pi \rho_s\int_0^{\infty}dr \frac r {1-f^2} 
[f' \pm \frac {Z(1-f^2)} r]^2 \ \ \ ,
\end{equation}
we find the Bogomol'nyi bound on the energy
\begin{equation}
\label{bogomolnyi}
E_G[f] \ge 4 \pi \rho_s |Z| \ \ \ ,
\end{equation}
and solving $rf' \pm Z(1-f^2) =0$ we obtain the BP solutions,
\begin{equation}
\label{sigma solution}
f(r) = \frac {(r/\lambda)^{2|Z|}-4} {(r/\lambda)^{2|Z|}+4} \ \ \ ,
\end{equation}
with energy $E_G = 4 \pi \rho_s |Z|$.

The energy $E$ is independent of the sign of $Z$, hence each solution $f(r)$ to 
(\ref{DE1})
describes both a skyrmion ($Z > 0$) and an antiskyrmion ($Z<0$). In the following, 
when we talk about skyrmions, and assume $Z$ is positive, the results are 
equally valid for antiskyrmions with negative charge provided we replace
$Z$ by $|Z|$. 

By contrast to the BP solutions, the skyrmions in the 
QH ferromagnet have a definite size
set by a competition between the Coulomb energy, $E_C$, and the Zeeman 
energy, $E_{\cal{Z}}$. 
The Coulomb energy favors large skyrmions since it decreases when 
charge is spread out. The Zeeman energy, on the other hand,  
increases when more spins are flipped, {\it i.e.} when the 
skyrmion becomes larger, and thus favors small skyrmions.
This leads to a competition that determines an optimal size and energy
of the skyrmion. 

In the next section we will construct approximate analytic solutions to
Eq. (\ref{DE1}) and in sections \ref{sec_relaxation} and \ref{sec_results} we will 
describe numerical solutions obtained
by a relaxation procedure.

\section{Analytic treatment}
\label{sec analytic}

We begin by recalling the Bogomol'nyi bound (\ref{bogomolnyi}) on the gradient 
energy of a configuration with topological charge $Z$, $E_G[f] \ge 4 
\pi \rho_s Z$. The Coulomb self-interaction of a charge distribution is 
non-negative, $E_C \ge 0$, and upon
dropping a negative extensive constant which implies that we are measuring 
excitation energies relative to the ferromagnetic ground state configuration,
$E_{\cal{Z}} \ge 0$ as well. Consequently, $E[f] \ge 4 \pi \rho_s Z$.

We will be interested in large skyrmions, for which the long-wavelength 
effective action approach will be accurate. As remarked previously, this
will be the case when $E_C$ ``dominates'' $E_{\cal{Z}}$, {\it i.e.} as 
$\tilde g \rightarrow 0$. In this limit of divergent skyrmion size, 
it is clear that $E_C \ll 1$
whence by a general balance argument the {\em value} of $E_{\cal{Z}} \sim E_C 
\ll1$
as well. Consequently, as $\tilde g \rightarrow 0$, the energy of the skyrmion
will approach the Bogomol'nyi bound and we expect that the solution itself
will converge to a BP solution with an appropriately chosen $\lambda$.

It turns out that this observation is sufficient to determine the leading small
$\tilde g$ characteristics of the skyrmions with $Z \ge 2$, but needs to be
supplemented by global considerations for $Z=1$. These follow from identifying
the various length scales in the problem by pairwise balancing the  
terms, $E_G \sim 1, \, E_C \sim 1/\lambda  \, {\rm and} \, E_{\cal{Z}} \sim 
\tilde g \lambda^2$, in the energy functional. This yields three scales: 
$R_1 \sim  1$ ($E_G$ and $E_C$), $R_2 \sim 1/\tilde g^{1/3}$ ($E_{\cal{Z}}$ and 
$E_C$) and 
$R_3 \sim 1/\tilde g^{1/2}$
($E_{\cal{Z}}$ and $E_G$). Note that {\em two} of these scales diverge as $\tilde g
\rightarrow 0$ with $R_3 \gg R_2$. Hence a global solution to (12) in this
limit can be expected to exhibit {\em both} of these scales.

We have not succeeded in constructing such a global solution, which appears
to be a difficult task due to the non-locality of the equation. 
Instead we will attempt to formally perturb around the BP solutions. To lowest 
order, this
is a problem in degenerate perturbation theory and requires merely that we
minimize $E_G + E_{\cal{Z}} + E_C$ in the subspace of the BP solutions. By 
construction,
$E_G$ is a constant in this subspace. The relevant
integral for the Zeeman term can be done analytically, and yields,
\begin{equation}
\label{zeeman2}
E_{\cal{Z}} = \nu_{{\scriptscriptstyle FM}} \pi \, \frac{2^{2/Z-1} 
\csc(\pi/Z)}{Z} \, \tilde g \lambda^2 \ \ \  .
\end{equation}
For $Z=1$, the double integral for the Coulomb term can be carried out 
as well:
\begin{equation}
E_C= \nu_{{\scriptscriptstyle FM}}^2 {3 \pi^2 \over 128} {1 \over \lambda} 
\ \ \ (Z=1) \ \ \ .
\end{equation}
However, for $Z \ge 2$ we had to take recourse to numerical integration with
the results,
\begin{eqnarray}
E_C &=& \nu_{{\scriptscriptstyle FM}}^2 {1.49 \over \lambda} \ \ \ (Z=2) \nonumber 
\\
&=& \nu_{{\scriptscriptstyle FM}}^2 {4.16 \over \lambda} \ \ \ (Z=3) \nonumber \\
&=& \nu_{{\scriptscriptstyle FM}}^2 {8.41 \over \lambda} \ \ \ (Z=4) \ \ \ .
\end{eqnarray}

Note that $E_{\cal{Z}}$ ($Z=1$) is infinite. Putting this aside for the moment, we
minimize $E_{\cal{Z}} + E_C$ with respect to $\lambda$ for $Z \ge 2$ to find,
\begin{eqnarray}
\label{charge234}
\lambda = 0.780 (\nu_{{\scriptscriptstyle FM}}/\tilde g)^{1/3}\ 
{\rm and} \nonumber \\
E = 2\bigl[ \sqrt{\pi \over 32} + 
1.43 &(&\nu_{{\scriptscriptstyle FM}}^{5}\tilde g)^{1/3} \bigr]\ \ (Z=2) \nonumber \\
\lambda = 1.29 (\nu_{{\scriptscriptstyle FM}}/\tilde g)^{1/3}\ {\rm and} \nonumber \\
 E = 3\bigl[ \sqrt{\pi \over 32} +
1.61 &(&\nu_{{\scriptscriptstyle FM}}^5 \tilde g)^{1/3} \bigr]\ \ (Z=3) \nonumber \\
\lambda = 1.75  (\nu_{{\scriptscriptstyle FM}}/\tilde g)^{1/3}\ {\rm and} \nonumber \\ 
E = 4\bigl[ \sqrt{\pi \over 32} + 
1.80 &(&\nu_{{\scriptscriptstyle FM}}^5 \tilde g)^{1/3} \bigr]\ \ (Z=4) \ \ \  .
\end{eqnarray}
We will see later on that these provide an excellent description of the
numerically determined solutions at small $\tilde g$, as expected.

The physically interesting case of $Z=1$ is more difficult. One can see
from Eq. (\ref{sigma solution}) that $1-f(r) \sim 1/r$ at large $r$, 
whence $E_{\cal{Z}}$ diverges
logarithmically with system size. Consequently, the BP solutions {\em cannot}
be used to globally approximate the true solutions at any value of $\tilde g
\ne 0$; the latter must exhibit a faster decay of $1-f(r)$ at large $r$.

The asymptotic behavior of the true solutions can be obtained directly from
the equation of motion (\ref{DE1}). At points far from the core, the Zeeman term
in the equation dominates over the Coulomb term. The latter is the potential 
of a localized charge distribution and will decay asymptotically as $1/r$. 
Dropping the latter, setting $f=1-\psi^2/2$ and linearizing
shows that $\psi$ satisfies Bessel's equation,
\begin{equation}
r^2 \psi'' + r \psi' - (Z^2 + {\nu_{{\scriptscriptstyle FM}} 
\tilde g r^2 \over 4 \pi \rho_s}) \psi = 0
\end{equation}
with the physical solution ($\psi \rightarrow 0$ as $r \rightarrow \infty$)
$\psi \propto K_Z(r/R_3)$ in terms of the scale $R_3= \sqrt{4 \pi \rho_s
\over \nu_{{\scriptscriptstyle FM}} \tilde g}$ introduced earlier. 
As $K_Z(x) \sim \sqrt{\pi \over 2x} e^{-x}$ for $x \gg 1$, the true asymptotic 
behavior of $1-f$ is exponential for {\em all} $Z$, different from the 
corresponding BP solutions.

For $Z \ge 2$, this discrepancy will (presumably) be attenuated by going
to higher orders in the expansion about the BP solutions - but it has no
bearing on the validity of the leading small $\tilde g$ behavior derived
previously. For $Z=1$, we will make the {\em assumption} that matching an
inner BP solution to an outer exponentially decaying solution (the minimal
fix for the divergence encountered) will yield the leading small $\tilde g$
behavior. (Such a solution will certainly converge pointwise to a BP solution,
but we have not shown that the corrections to it are subdominant. For $Z\ge2$
one can readily show that this is the case by linearizing about the optimized
BP solution \cite{abanov}.) More precisely, we assume that,
\begin{eqnarray}
\label{matched}
f(r) &=& { (r/\lambda)^2 - 4 \over (r/\lambda)^2 + 4} \ \ , \ \  r\ll R_3 
\nonumber \\
     &=& 1 - {a^2_Z \over 2} K_1^2 ({r \over R_3}) \ \ , \ \ r \gg \lambda
\end{eqnarray}
is the desired solution at small $\tilde g$ (for $Z=1$: $a^2_1 = 16 \lambda^2/R_3^2$).
It is easy to check that with this choice of $a$, both pieces of the definition
of $f$ agree on the interval $ \lambda \ll r \ll R_3$ where they overlap.

Assuming $\lambda \ll R_3$, we find that,  with the above choice of $f$, 
\begin{equation}
\label{zeeman1}
E_{\cal{Z}}= 4 \nu_{{\scriptscriptstyle FM}} \tilde g 
\lambda^2 \ln ({R_3\over  a_1 \lambda}) \ \ \ ,
\end{equation}
where $\ln(a_1) = {1 \over 2} + \gamma $, {\it i.e.}  $a_1 = 2.9365$. Combining this
with our earlier expression for $E_C$, we are required to minimize
\begin{equation}
E(\lambda) = {\beta \over \lambda} + \alpha \tilde g \lambda^2 \ln({1\over
\lambda} \sqrt{\delta\over\tilde g})
\end{equation}
with $\alpha = 4 \nu_{{\scriptscriptstyle FM}}$, $\beta = 
{3 \pi^2 \nu_{{\scriptscriptstyle FM}}^2\over128}$ and 
$\delta = 4 \pi \rho_s/a_1^2 
\nu_{{\scriptscriptstyle FM}}$ in the region $\lambda \ll R_3$, {\it i.e.} $\lambda 
\sqrt{\tilde g} \ll 1$.
Differentiating with respect to $\lambda$ and setting $\lambda^3 = 
(3 \beta/2 \alpha) \lambda'^3$ then leads to the transcendental equation,
\begin{equation}
\label{lambdatrans}
{1 \over \lambda'^3} = \tilde g \ln({c\over\lambda'^3 \tilde g^{3/2}})
\end{equation}
with $c=(2 \alpha/3 \beta) (\delta/e)^{3/2}$. As $c$ involves $\rho_s$, 
which depends on $\nu$, it does not scale trivially with $\nu$; 
for $\nu=1$ it equals $0.0178$. Eq. (\ref{lambdatrans}) has solutions only for 
$\tilde g$ below some critical value, $\tilde g_{c_3}$. For $\nu = 1$, one finds
$\tilde g_{c_3} = 4.3 \times 10^{-5}$. At this point $\lambda = 12.6$, hence the 
condition $\lambda \sqrt{\tilde g} \ll 1$ is obeyed and we also note that 
$c/\sqrt{\tilde g_{c_3}} = 2.7$.

Equation (\ref{lambdatrans}) can be solved perturbatively at small $\tilde g$ by replacing
$\lambda'^3$ in the argument of the logarithm by the full RHS. 
This yields, to second order in this procedure,
\begin{eqnarray}
\label{lambda1}
{1 \over \lambda^3} = {2 \alpha \over 3 \beta}\, \tilde g \, \{ \ln( c / \sqrt 
{\tilde g})
+\ln{\ln(c/\sqrt {\tilde g})} \nonumber \\
+ O( \ln{\ln(c/\sqrt {\tilde g})} / 
\ln(c/\sqrt {\tilde g}) ) \}
\end{eqnarray}
for the parameter $\lambda$ and thereafter
\begin{eqnarray}
\label{energy1}
E(\tilde g) = \sqrt{\frac {\pi} {32}}+A [\tilde g \ln(c/\sqrt {\tilde g})]^{1/3} 
\{ 1 + {1\over3} 
{\ln\ln(c/\sqrt {\tilde g}) \over \ln(c/\sqrt {\tilde g})} 
\nonumber \\
+ {1\over2} 
{1 \over \ln(c/\sqrt {\tilde g})} +O( (\ln\ln(c/\sqrt {\tilde g}) / 
\ln(c/\sqrt {\tilde g}))^2 ) \}
\end{eqnarray}
(with $A= 3* (\alpha/3)^{1/3} (\beta/2)^{2/3}=0.7838$) for the energy of the 
skyrmion. In all of these expressions, the scale for $\tilde g$ is set
by $c^2 \approx 3 * 10^{-4}$ which is quite small. This comes about
as the argument of the logarithm is essentially the ratio $R_3/R_2$
and the asymptotic regime starts to make sense only when this ratio is
$O(1)$. As this ratio grows extremely weakly, as $\tilde g^{1/6}$, one
has to go to fairly small $\tilde g$ before entering asymptopia and as
a result the  scale of the logarithms, which is set by the scale of this 
crossover, is small. We also note, as discussed at more length elsewhere,
that the logarithmic enhancement of the energies of the $Z=1$ skyrmions
relative to their $Z=2$ cousins causes them to bind at extremely small
values of $\tilde g$ \cite{lilliehook}.

Finally, we can use the leading solutions developed here to compute various
measures of the size of the skyrmions. Quantities of particular interest
are the spin of the skyrmions and their root mean squared spin and charge
radii defined via:
\begin{eqnarray}
\label{radii}
S_z &=&  \overline{\rho} \int d^2r \frac 1 2 (1-f(r)) \nonumber \\
R_s^2 &=&\langle r^2 \rangle _s \equiv \overline{\rho} \int d^2r 
r^2 \frac 1 2 (1-f(r)) /  S_z  \nonumber \\
R_c^2 &=& \langle r^2 \rangle _c \equiv {1\over Z} \int d^2r r^2 q(r) 
\ \ \ ,
\end{eqnarray}
where we have approximated the local density of electrons, $\rho(r) = 
\overline{\rho} + \nu_{\scriptscriptstyle FM} q(r)$, 
by $\overline{\rho} = \nu_{{\scriptscriptstyle FM}}/2\pi$. 
These quantities are not all independent; an integration by parts shows that 
$S_z = \nu_{{\scriptscriptstyle FM}} R_c^2/2$, and, moreover, since 
$E_{\cal{Z}}=\tilde g S_z$, the spin is obtained from (\ref{zeeman2}) and 
(\ref{zeeman1}).

%For the spin we find,
%\begin{eqnarray}
%S_z &=& 4 \nu_{{\scriptscriptstyle FM}} \lambda^2 \ln({R_3\over a_1 \lambda}) 
%\ \ \ 
%(Z=1) \nonumber \\
%    &=& {\pi\over2} \nu_{{\scriptscriptstyle FM}} \lambda^2 \ \ \ (Z=2)
%\end{eqnarray}
For the spin radius we find,
\begin{eqnarray}
R_s &=& \sqrt{2 \over 3} 
{R_3 \over \ln^{1/2}({R_3\over a_1 \lambda})} 
\ \ \  (Z=1) \nonumber \\
    &=& 2\sqrt{2 \over \pi} \lambda \ln^{1/2}({R_3\over a_2 \lambda}) \ \ \ (Z=2) 
\nonumber \\
    &=& (2^{2/Z-1} \sec (\pi/Z) )^{1/2} \lambda \ \ \ (Z \ge 3)
\end{eqnarray}
with $\ln(a_2) = \gamma - {1\over12}- {\ln 2\over 2}$.  Note the feature, that
depending on the charge of the skyrmion and the highest power of radius in
the integrand, we get different scales entering the leading dependence.

\section{Numerical methods}
\label{sec_relaxation}
Here we describe how we numerically solve Eq. (\ref{DE1}).
We use mainly a relaxation technique \cite{relaxation},  however, due to numerical 
problems at $r=0$ for $\tilde{g}  > 10^{-3}$, we must combine this 
with a shooting method in order to achieve high
numerical precision \cite{shoot}. We begin by describing the relaxation method.

\subsection{Relaxation technique}
The idea of the relaxation method
is to start from an ansatz, $f$,  for a solution to the
differential equation (DE),  ${\bf D}(f)=0$,
and then
use a first order Taylor expansion to improve this ansatz
iteratively. We want to find $\Delta f$ so that 
${\bf D}(f + \Delta f)=0$; expanding to first order we have
\begin{equation}
\label{taylorexp}
 {\bf D}(f) + \int dr \frac{\delta {\bf D}}
{\delta f}\cdot \Delta f = 0 \ \ \ .
\end{equation}
Solving for  $\Delta f$ gives the 
improved function $f+\Delta f$.
To implement this on a computer, we first
rewrite the $N^{\rm th}$ order 
differential equation as $N$ coupled 
first order equations (in our case $N=2$). We then put the  system on 
a lattice and replace the differential equations by
finite difference equations (FDE). The resulting algebraic 
equations relate the values of the functions ${\bf f} = (f^1,f^2,...,f^N)$ 
(where $f^1=f, \, f^2=df/dr \ldots$)
at lattice points $k$ (${\bf f}_k$) and $k+1$ (${\bf f}_{k+1}$):
\begin{equation}
\label{eq: relax}
{\bf D}_k(r_k,r_{k+1},{\bf f}_k,{\bf f}_{k+1}) = 0 \ \ \ ,
\end{equation}
here ${\bf D}_k$ are the FDEs at lattice point $k$. 
Taylor expanding as in (\ref{taylorexp}) gives
\begin{eqnarray}
{\bf D}_k({\bf f}_k + \Delta {\bf f}_k, 
{\bf f}_{k+1} + \Delta {\bf f}_{k+1}) 
\approx {\bf D}_k({\bf f}_k,{\bf f}_{k+1}) 
\nonumber \\
+ \sum_{\alpha=1}^N \frac{\partial {\bf D}_k}
{\partial f^{\alpha}_{k}} \Delta f^{\alpha}_{k} + 
\sum_{\alpha=1}^N \frac{\partial {\bf D}_{k}}
{\partial f^{\alpha}_{k+1}} \Delta f^{\alpha}_{k+1} = 0 \ \ \ .
\end{eqnarray}
These equations are then solved for
the increments $\Delta {\bf f}_k$ and the procedure 
is iterated until the
corrections, $\Delta{\bf f}_k$, are small enough. As a 
measure of the error we use:
\begin{equation}
err = \frac{1}{M \cdot N} \sum_{k=1}^{M}\sum_{\alpha=1}^{N} | \Delta
f^\alpha_k | \ \ \ ,
\end{equation}
where $M$ is the number of lattice points \cite{error}.
If the 
error is large only a fraction of the correction 
is used when calculating the new function, this
reduces the risk for overcorrection near $f=\pm 1$
where the DE is very sensitive to disturbances in $f$.
After each iteration
we calculate the energy, using 
equation (\ref{eq: total energy}), and check
that it  decreases. 

%\begin{eqnarray}
%ansatz \rightarrow Coulomb\rightarrow 
%relax \rightarrow Coulomb 
%\rightarrow relax ... \nonumber
%\end{eqnarray}

\subsection{Numerical solution for small $\tilde{g}$}

There are two problems to solve before we can apply the relaxation method: 
First, the boundary conditions are given
at $r=0$ and at $r=\infty$ but we can only cover a finite interval
$[0,r_{max}]$ with the $M$ lattice points  (with the DE 
in the above form). This is a problem for large skyrmions, {\it i.e.} 
when $\tilde g$ is small. Second, 
the relaxation method relies on the validity 
of a first order Taylor expansion, therefore it is crucial
to make an ansatz that is close to the true solution.

\begin{figure}
\begin{center}
\leavevmode
\psfig{file=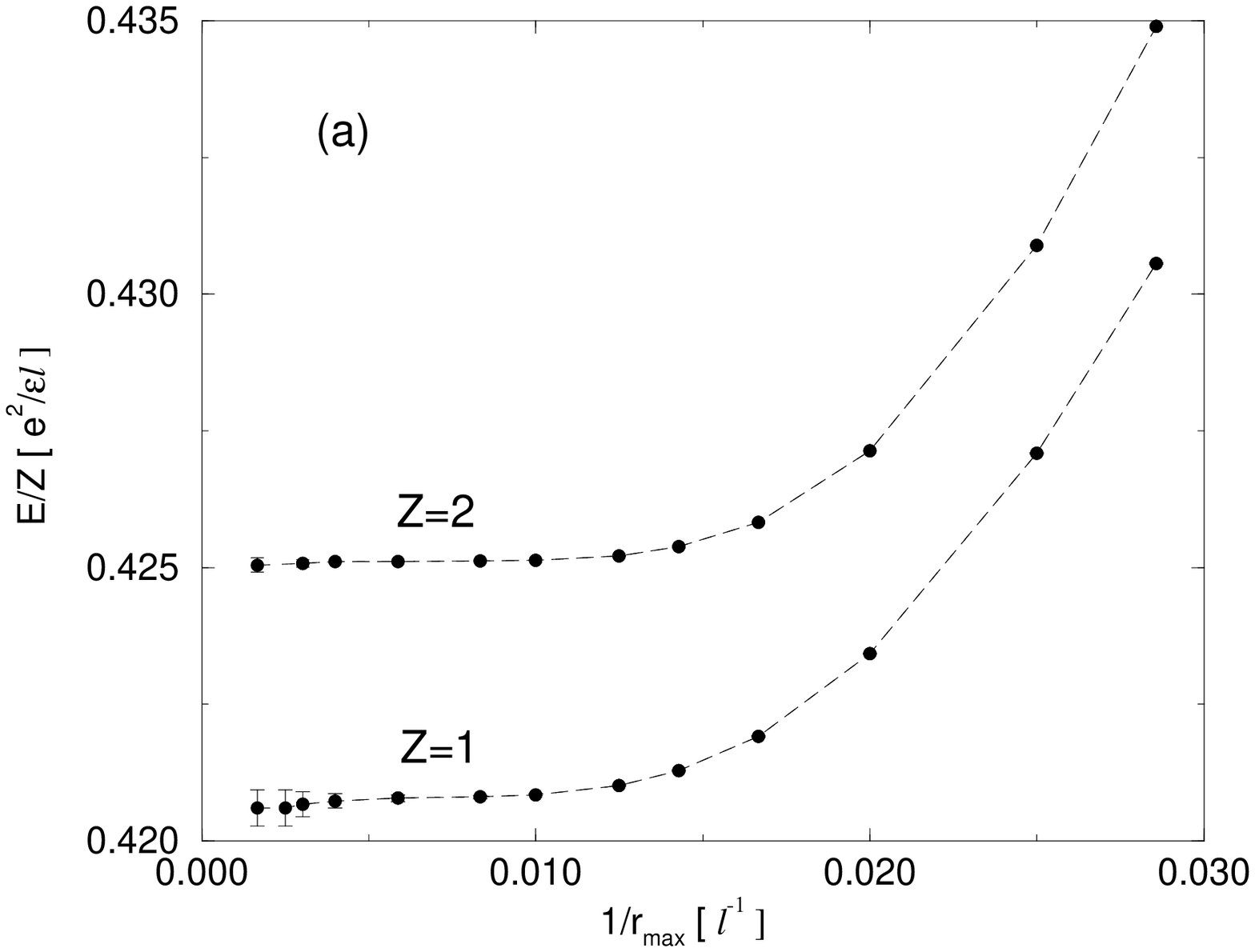,height=7cm}
\psfig{file=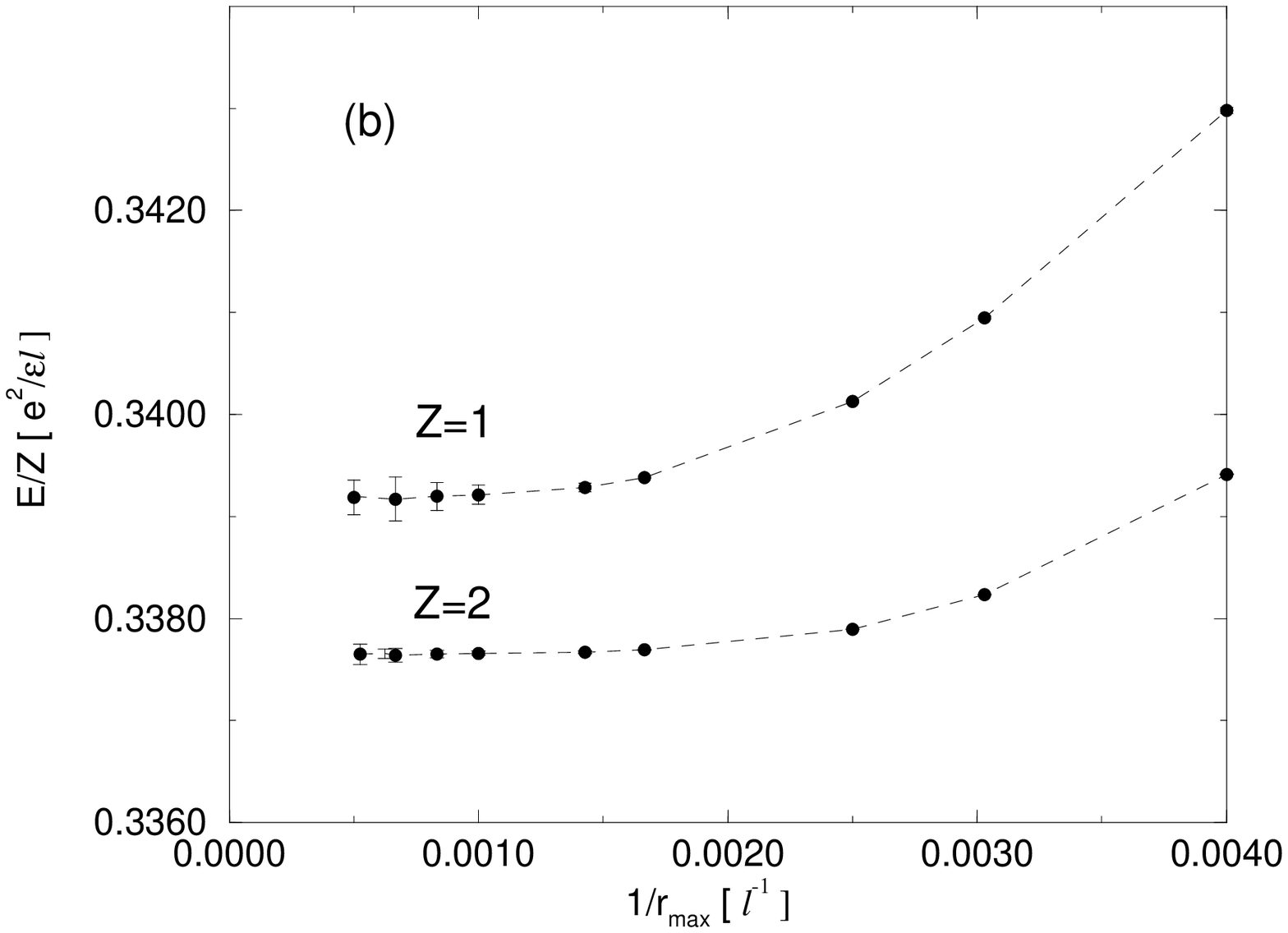,height=7cm}
\end{center}
\caption{Skyrmion energies (per charge $Z$) as functions of system size for
(a) $\tilde{g}=5 \cdot 10^{-4}$ and (b) $\tilde{g}=5 \cdot
10^{-6}$, at $\nu=1$.}
\label{fig: E scaling}
\end{figure}

To minimize the finite size effects we solve the equations 
for several different $r_{max}$ and extrapolate  to infinite system size.
Fig. \ref{fig: E scaling} shows how the energy scales with $1/r_{max}$ 
for $\tilde{g}=5 \cdot 10^{-4}$ and $5 \cdot 10^{-6}$. We see that the 
energy is more or less independent of system size at the largest sizes, whence
the finite size effects are negligible. The spin, and the spin radius, 
show bigger finite size effects and scaling to infinite system size is necessary 
to obtain accurate values when $\tilde g \lesssim 10^{-6}$.  

To obtain a good initial ansatz is a bit tricky; we 
proceed as follows.
The Belavin-Polyakov solution, Eq. (\ref{sigma solution}), to the scale invariant 
model 
is expected to be a good
approximation to the true solution in 
the core region (for some value of the scale parameter $\lambda$), 
see Sec. \ref{sec analytic}.
As our ansatz we therefore 
use this solution
with the function $const \cdot r^{Z}$ 
added, where we choose the constant so that the 
boundary condition $f(r_{max})=1$ is fulfilled.
We run the relaxation
procedure for different $\lambda$ and choose the one for
which the convergence is best.

Having determined an ansatz we can start iterating.
The Coulomb term $V_C(r,f')$, 
which enters the DE, depends on the integral 
over $f'$. Thus we start each  iteration by calculating the function
$V_C(r,f')$ using the most recent $f$.
Then we iterate until the error is small enough,
each time checking that the energy decreases.

To check the accuracy of the solution we use a shooting test.
The relaxation solution is used as input to the
shooting method (fourth order Runge-Kutta). We pick a
point $r_0$ close to $r=0$ ({\it e.g.} the first lattice 
point where $f'\ne 0$) and treat the problem as 
an initial value problem with $f(r_0)$ and $f'(r_0)$ given
from the relaxation solution. If the shooting 
solution thus obtained coincides with the relaxation solution 
we are confident that this is the right solution.
Comparing the energy obtained from the relaxation
and the shooting solutions we find, for both $Z=1$ and $Z=2$, that  for the smallest 
$\tilde{g}$ ($\tilde g \approx 10^{-7}$) this difference
is  of the order of $10^{-5}$. The difference is approximately
constant up to $\tilde{g}=10^{-4}-10^{-3}$.
For $\tilde{g} \gtrsim 10^{-3}$ the difference starts to grow; 
this is also reflected in the convergence properties of the 
relaxation method.
This indicates that the ansatz we use deviates more from the 
true solution and the relaxation method has problems avoiding
the singular behavior near $r=0$.

\subsection{Larger $\tilde{g}$} 

\begin{figure}
\begin{center}
\leavevmode
\psfig{file=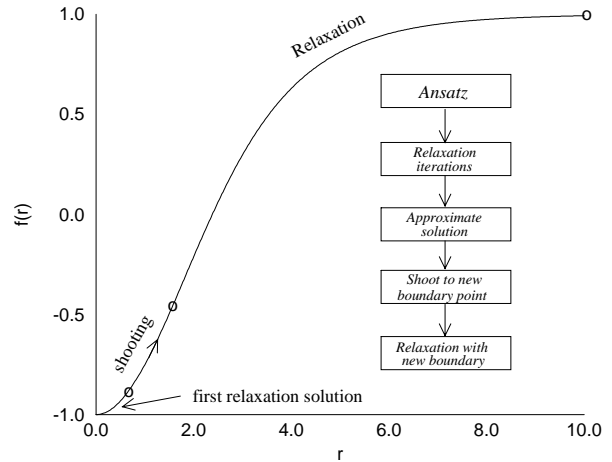,height=7cm}
\end{center}
\caption{Sketch of the numerical method used to
solve the equations of motion for $\tilde g > 10^{-3}$.}
\label{fig: solution method}
\end{figure}

The problems for larger $\tilde{g}$ ($\tilde g \gtrsim  10^{-3}$)
has to do with the structure of the
differential equation (\ref{DE1}) at the origin 
together with a smaller overlap between the solution and the
ansatz. In this region we proceed as follows.
The solution obtained from an initial set of relaxation iterations
is used to shoot from the point $r_0$ \cite{shoot}
(close to $r=0$) to a point
a finite distance away from $r=0$.
This point is then used as 
a boundary point for a set of new relaxation iterations.
With this method, which is illustrated 
in Fig. \ref{fig: solution method}, we avoid the problems near $r=0$
\cite{alt}.

This method certainly introduces an error in the 
calculation. As a rough estimate of the error we take 
the energy difference between the initial relaxation solution
and the solution obtained by the 
combined relaxation and shooting method.
For $\tilde{g}=0.02$ this is of the order of one percent.
(Hartree-Fock  theory\cite{fertig}
shows that the skyrmion 
has a size of a few magnetic lengths and 
consists of a few flipped spins at  $\tilde{g}=0.02$. Whereas Hartree-Fock
should be correct in this region, the validity of the effective 
long-distance theory becomes questionable at distances of the order of the
magnetic length.) 
For a  comparison between Hartree-Fock and effective 
action results see Sec. \ref{sec_results} below.

\section{Numerical results}
\label{sec_results}

\begin{figure}
\begin{center}
\leavevmode
\psfig{file=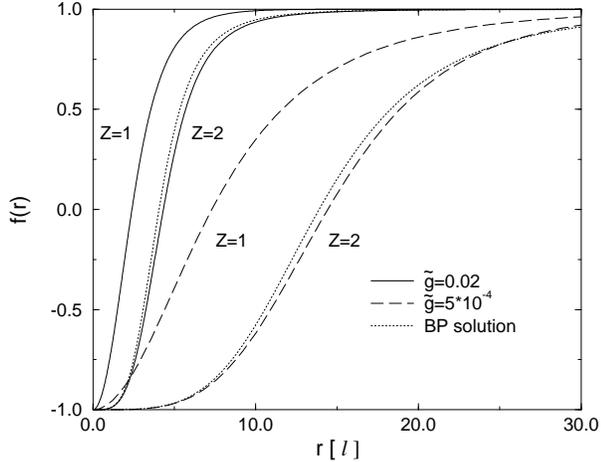,height=7cm}
\end{center}
\caption{The relaxation solution, $f(r)$, for
${\tilde g}=0.02,\, 0.0005$ and $Z=1,2$. $Z=2$ is  compared to the 
Belavin-Polyakov 
solution (\ref{sigma solution}),
with the scale, $\lambda$, taken from Eq. (\ref{charge234}).}
\label{fig: f}
\end{figure}

\begin{figure}
\begin{center}
\leavevmode
\psfig{file=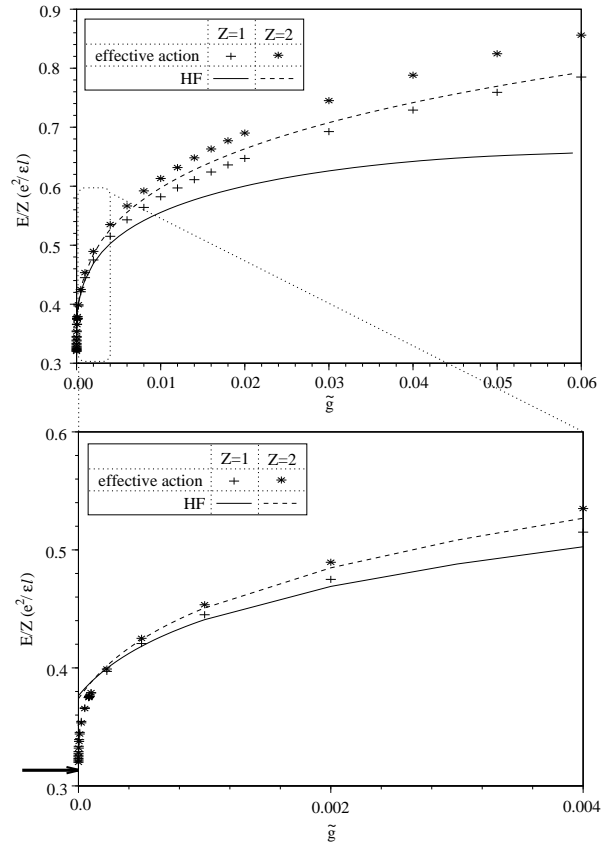,width=18cm}
\end{center}
\caption{Effective action and Hartree-Fock results for 
the energy, $E$, per charge (relative to the ground state)
 of the  $Z=1,2$ skyrmions as function of $\tilde{g}$.
(Due to finite size effects in the HF calculation the HF energy 
becomes larger than the effective action energy for very small $\tilde{g}$.) 
The arrow in the inset marks the energy at $\tilde g =0$.}
\label{fig: energy}
\end{figure}

\begin{figure}
\begin{center}
\leavevmode
\psfig{file=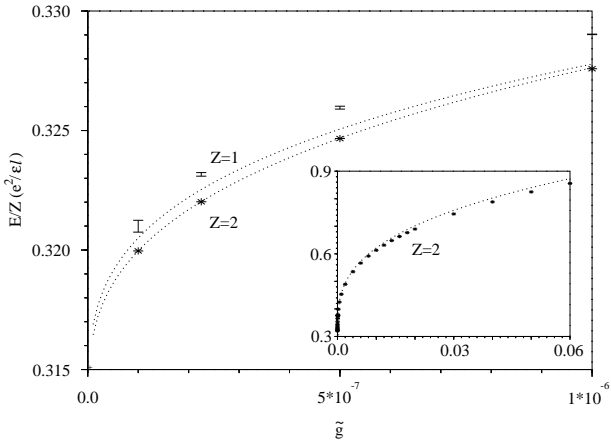,height=6.5cm}
\end{center}
\caption{Numerical effective action results (points) for the energy
compared with the analytic 
expressions, Eq. (\ref{energy1}) and (\ref{charge234}), 
(lines) for $Z=1,2$ skyrmions. The error 
bars come from finite size effects, for $Z=2$ the error
bars are smaller than the dots. The inset shows the $Z=2$ 
data for large $\tilde g$.}
\label{fig: E_ana1}
\end{figure}

\begin{figure}
\begin{center}
\leavevmode
\psfig{file=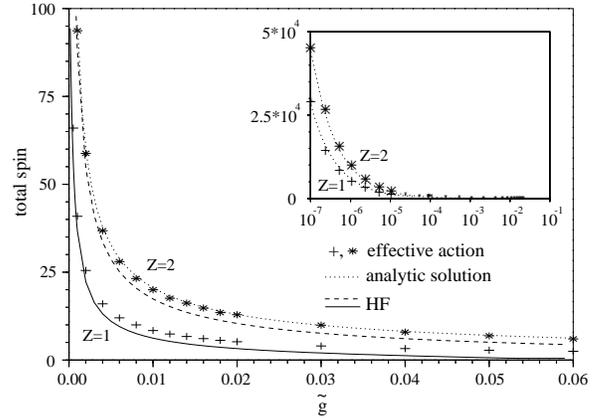,width=9cm}
\end{center}
\caption{Spin, $S_z$, for $Z=1,2$ skyrmions;  numerical effective action results 
compared to HF and analytic small $\tilde g$ expansions.
The inset shows results for very 
small $\tilde{g}$ where no HF data are available. For $Z=1$, the analytic expansion 
breaks down at $\tilde g_{c_3}=4.3 \times 10^{-5}$. }
\label{fig: spin}
\end{figure}

\begin{figure}
\begin{center}
\leavevmode
\psfig{file=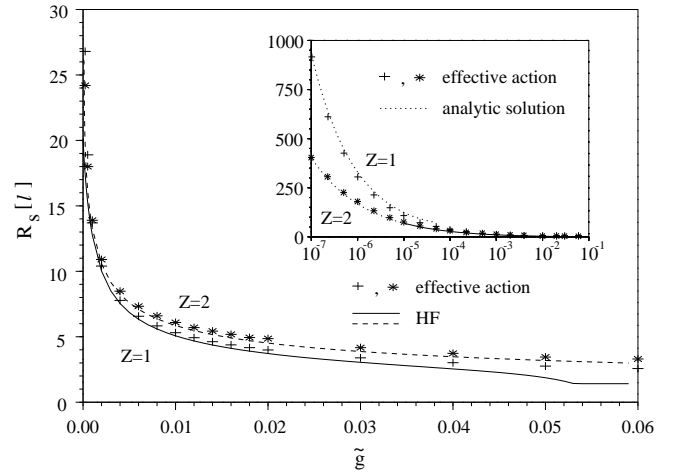,height=7cm}
\end{center}
\caption{Spin radii for $Z=1,2$ skyrmions; numerical effective action results
compared to HF and analytic small $\tilde g$ expansions. 
The spin radii for
$Z=1$ and $Z=2$ are equal at $\tilde g \sim 10^{-3}$.} 
\label{fig: radius}
\end{figure}

\begin{figure}
\begin{center}
\leavevmode
\psfig{file=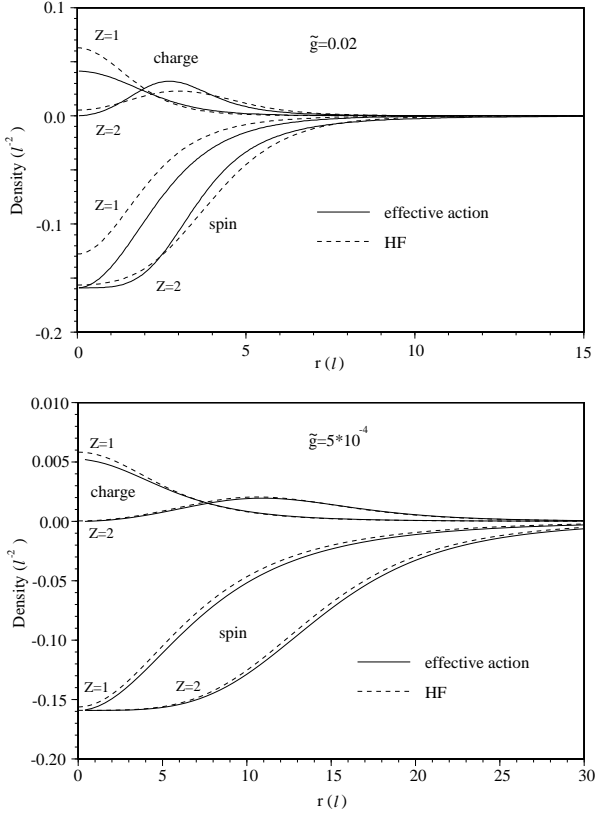,width=9cm}
\end{center}
\caption{Spin and charge density profiles for $Z=1,2$ skyrmions; effective action 
and 
HF results.}
\label{fig: densities}
\end{figure}

We have applied the technique described in Sec. \ref{sec_relaxation} to skyrmions 
with
topological charge $Z=1,2$ and $3$ at $\nu=1$, {\it i.e.} 
$\nu_{{\scriptscriptstyle FM}}=1$, 
$\rho_s = 1/16\sqrt{2\pi}$. Our results 
are presented  in Figs. \ref{fig: f}-\ref{fig: densities}. The $Z=3$ 
skyrmion is not found to be the 
lowest energy excitation for any $\tilde{g}$ considered here (in
agreement with the analytic result, Sec. \ref{sec analytic}), we therefore
show results for $Z=1$ and $Z=2$ skyrmions only.

Fig. \ref{fig: f} shows the $z$-component of the spin, $n_z=f$, as a function 
of the radius, $r$. An  analysis shows that for $Z=2$ at $\tilde 
g = 5\cdot 10^{-4}$, 
the numerical solution
agrees very well in the core region with the Belavin-Polyakov solution 
(\ref{sigma solution}),  with $\lambda$ determined
by Eq. (\ref{charge234}). For $Z=1$, no analytic solution is available 
for the values 
of $\tilde g$ shown in the figure, however for   
$\tilde g < \tilde g _{c_3} = 4.3 \cdot 10^{-5}$,
the numerical solution agrees  well with the analytic solution 
(\ref{sigma solution}),  
with $\lambda$ determined
by Eq. (\ref{lambda1}).
Away from the core region, the  relaxation 
solutions differ  from the BP solutions, although the differences are 
quite small for $Z=2$.

From the numerical solution $f(r)$ we calculate the energy of the skyrmion 
using Eq. (\ref{eq: total energy}). In Fig. \ref{fig: energy}, the
energy of the  $Z=1,2$ skyrmions, relative to the groundstate energy 
$E_{(f=1)}=-\frac 1 4 {\tilde g} r^2_{max}$, is given as a function of $\tilde g$ 
and compared to Hartree-Fock (HF) results \cite{fertig,enconv}. We see
that the effective  and Hartree-Fock theories
agree very well for small $\tilde g$, 
but that the HF energy is substantially lower  for larger values 
of $\tilde g$. In particular, the effective 
theory predicts that the skyrmions are the lowest energy charged excitations 
for $\tilde{g} < 0.018$ whereas HF gives  $\tilde{g} < \tilde g _{c_1} = 0.054$. 
The difference between 
effective action and Hartree-Fock results is smaller for $Z=2$ than for $Z=1$.
The effective theory is a long distance theory valid only for large smooth
skyrmions, {\it i.e.} for small enough $\tilde g$. Hartree-Fock theory, on
the other hand, is valid all the way from polarized quasiparticles and  
very small skyrmions near $\tilde g_{c_1}=0.054$ to quite large skyrmions (where 
finite size effects eventually limit the applicability of HF). 
It is interesting to note that the effective theory 
works quite well also for 
fairly large values of $\tilde g$. 
In the limit of very small $\tilde g$, the 
effective theory becomes exact (see Sec. \ref{sec analytic}) and $E/Z$ approaches 
$\sqrt{\pi /32}$ (marked by $\rightarrow$ in the figure). The blowup in
Fig. \ref{fig: energy} shows a region (close to $\tilde{g}=0$) where 
the Hartree-Fock energy is larger than the effective action energy, this
is an effect of the difficulty of handling large enough systems in 
Hartree-Fock. Since in this range, the 
skyrmions become very large (see Fig. \ref{fig: radius}) finite size 
effects make the Hartree-Fock energies unreliable.
For $\tilde g< \tilde g_{c_2}=8.4 \times 10^{-5}$, the $Z=2$ skyrmions
actually have lower energy (per charge) than the $Z=1$ 
skyrmions, {\it i.e.} the $Z=1$ skyrmions bind in pairs \cite{lilliehook}.

We now  compare the analytic small $\tilde{g}$
expansions for the energy, derived in 
Sec. \ref{sec analytic}, with the numerical effective action results.
Fig. \ref{fig: E_ana1} shows the skyrmion energies for 
$10^{-7} \lesssim \tilde{g} \lesssim 10^{-6}$; the inset shows the $Z=2$ energies
for larger $\tilde g$. 
For $Z=2$ the agreement is excellent all the way up to $\tilde g =0.06$.
 For $Z=1$, the numerical results are seen to approach the analytic 
result  Eq. (\ref{energy1}) as $\tilde g \rightarrow 0$, however, deviations 
are still visible at $\tilde{g}=10^{-7}$. The size of these is consistent with the 
size of the ignored correction terms in Eq. (\ref{energy1}). Note that the correction 
terms decrease extremely slowly with decreasing $\tilde g$. 

Fig. \ref{fig: spin} shows the spin, $S_z$, of the $Z=1,2$ skyrmions relative to the 
groundstate,  see Eq. (\ref{radii}). Since  $S_z=R_c^2/2$ this also gives the 
(topological) charge radius, $R_c$. The numerical effective action results 
and the Hartree-Fock results become indistinguishable, for $Z=1$ as well as for
$Z=2$, as 
$\tilde g$ approaches zero but, in fact, agree quite well over 
the whole range although deviations are observable for larger $\tilde g$. The analytic
small $\tilde g$ expansions agree extremely well with the numerical effective action 
results for all $\tilde g$ for $Z=2$ and for $\tilde g < \tilde g _{c_3}$ for $Z=1$.

Fig. \ref{fig: radius} shows how the spin radii, $R_s$,
see Eq. (\ref{radii}), of the skyrmions decrease with increasing $\tilde g$.
The $Z=2$ spin radius is smaller than the $Z=1$ radius for 
$\tilde g \lesssim 10^{-3}$. 

In  Fig. \ref{fig: densities} we show spin and charge density profiles. 
A comparison between the
Hartree-Fock and effective action densities reveals
almost identical profiles at $\tilde{g}=5 \cdot 10^{-4}$, see
\cite{deviation}, whereas at $\tilde{g}=0.02$ the profiles
deviate substantially. We also note 
that the agreement for the $Z=2$ skyrmion is better 
than the agreement for the $Z=1$ skyrmion  (at large $\tilde{g})$.

\section{Asymptotic shape of the skyrmion}
\label{sec_asy}

Here we will examine the detailed profile of the skyrmions in two
regions, near the center and far from the center, where it is possible to 
compute it rigorously. This serves as a further check on our analysis.

\subsection{Small $r$}

Since the charge distribution is radially symmetric
it is clear that $\partial V_C / \partial r |_{r=0}=0$.
%a calculation shows that 
%$\partial V_C / \partial r \propto r/\lambda^3$
%for $r \ll \lambda$ ($Z=1$).
This implies that both the Zeeman and the Coulomb terms
in the differential equation (\ref{DE1}) can be neglected  and 
the solution is given by the Belavin-Polyakov
solution, Eq. (\ref{sigma solution}).
This is consistent with the numerical solutions, see Fig. 
\ref{fig: f}.

\subsection{Large $r$}

\begin{figure}
\begin{center}
\leavevmode
\psfig{file=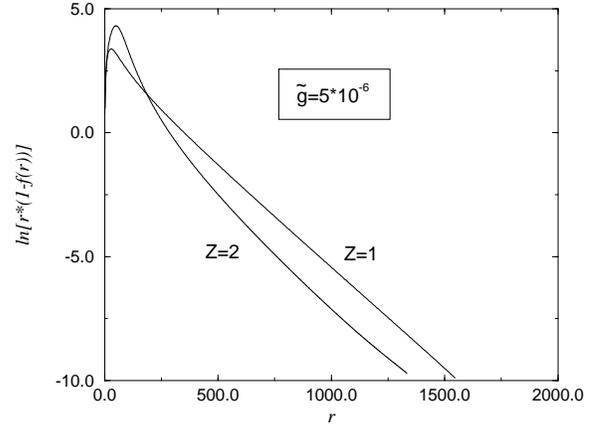,height=6.5cm}
\end{center}
\caption{
Numerical effective action result
for the asymptotic behavior of
$f(r)$, at $\nu=1$. Both 
the $Z=1$ and the $Z=2$ cases are well described
by the asymptotic formula derived in the text.}
\label{fig: asymptotic f}
\end{figure}

As we showed in Sec. \ref{sec analytic}, far from the core, $r \gg R_3$, the
skyrmion profile, $1-f(r) \propto K^2_Z(r/R_3)$, is exponential, 
\begin{equation}
\label{asymptot Bessel}
1-f(r) \propto {R_3 \over r} e^{-2r/R_3} \ \ \ .
\end{equation}
The crossover to this regime can be estimated
by examining the behavior of the Bessel functions for {\em small}
values of their argument: 
\begin{equation}
K_Z(x) \sim {2^{Z-1} \Gamma(Z) \over x^Z} \ \ \ ,
\end{equation}
which shows that the crossover
 sets in at $x^Z \sim 2^{Z-1} \Gamma (Z)$, for example, at
$r\sim R_3$ ($Z=1$) and
$r \sim \sqrt{2} R_3$ ($Z=2$), {\it i.e.} farther out for higher $Z$.
To compare these observations with the numerical solution we plot 
$\ln[r (1-f(r))]$ against $r$ for $Z=1$ and $Z=2$, at $\nu=1$, 
see Fig. \ref{fig: asymptotic f}. 
When $\tilde{g}=5 \cdot 10^{-6}$, $R_3=250$ and  we get a straight line
for $r>400$ when $Z=1$ and for $r>700$ when $Z=2$; 
in qualitative agreement with our above observation.
Reading off the slope for $Z=1$  in the figure and comparing to
Eq. (\ref{asymptot Bessel}) one finds $R_3 = 247 \pm 9$, in 
excellent agreement with the exact value $R_3=250$. 
For $Z=2$, the exponential region sets in further out and, in addition, 
the  curve bends up slightly for
$r>1000$ due to finite size effects in combination with 
the finite resolution of real numbers in the computer,  
therefore this curve is harder to analyze. However,  comparing the curve
for several different system sizes and keeping the part 
which is unchanged, one finds that it is consistent with an exponential 
behavior with $R_3 =250$.

\section{Edges}
\label{sec_edg}
In previous work \cite{karlhede} we have used the effective theory 
to study the reconstruction of a sharp polarized edge of a semi-infinite
quantum Hall ferromagnet as the confining potential is softened. The
problem of reconstruction is as follows. 
For hard confinement, the ferromagnetic bulk state continues all the way 
out to the edge and the density drops abruptly to zero over the scale of 
the magnetic length.
When the edge potential softens, charge eventually moves out, {\it i.e.}
the edge reconstructs. The question is how this happens given
the competition between the correlation energy of the quantum Hall
state and the Hartree energy due to the confinement. 
In \cite{karlhede} we argued that the initial instability, as the potential 
softens, is to forming a spin texture along the edge\cite{others}.
Charge is then moved out as a consequence of the spin-charge relation of spin 
textures. Here we
will give details, previously unpublished, of how the formation of edge 
textures is analyzed within the effective theory. 
(Before proceeding, the following caveat is in order. The use of an
unmodified magnetic action near the edge is not unproblematic given that
there are gapless density modes near the edge among other things. While
we have not attempted to derive an action for the edge from microscopics,
we have checked \cite{karlhede} by comparison with Hartree-Fock calculations,
that the action we use is accurate for getting static properties of the
edge at small $\tilde g$.)

We consider a semi-infinite Hall system occupying the region $x<0$ ($y$ is the 
direction along the edge). As an ansatz for the spin vector field 
${\bf n}({\bf r})$ we take
\begin{eqnarray}
\label{eq: edge ansatz}
  n_x &=& \sqrt{1 - f^2(x)} \cos(ky) \ , 
\nonumber \\
    n_y &=& \sqrt{1 - f^2(x)} \sin(ky) \ , 
\nonumber \\
    n_z &=& f(x) \ . 
\end{eqnarray}
This leads to the topological density
\begin{equation}
\label{eq: edge charge density}
q(x) = \frac{k}{4 \pi} f'(x) \ \ \ ,
\end{equation}
where $f'(x)=df/dx$. The $z-$component
of the spin, $f(x)$, gradually falls from one in the bulk,
to some value $f_e$ ($-1\leq f_e\leq 1)$ at the edge. As one moves along the edge 
the spin in the $xy$-plane rotates with wave number $k$. $f_e$ and $k$ are 
continuous parameters that are to be determined. 
Note that, although 
locally this texture is identical to the texture for bulk skyrmions there is no 
quantization: $\int d^2r q({\bf r})$ can take any value. 

The electron density, $\rho(x)$, is 
\begin{eqnarray}
\label{rho}
 \rho  &=& \overline{\rho}+\nu_{{\scriptscriptstyle FM}}q(x) \ \ ,
\; \; \; \; \; -\infty <x\leq x_R \\
\nonumber
    \rho &=& 0 \ \ , \; \; \; \; \; \; \; \; \; \; \; x_R<x \ \ .
\end{eqnarray}
The unreconstructed, sharp polarized edge corresponds to taking 
$f(x)=1, \, x\leq 0$ and $f(x)=0, \, 0<x$. This case is obtained by 
letting $q(x)=0, \, x_R=0$ in  
(\ref{rho}). When the edge is textured, charge is moved out and $x_R$ is 
determined by charge conservation: $\int_{-\infty}^{x_R} dx \rho = 
\int_{-\infty}^{0} dx \overline{\rho}$. This gives,  
\begin{equation}
\label{eq: charge neutrality}
x_R = \frac{k}{2} (1-f_e)
 \ \ \ .
\end{equation}

We assume the confining potential is caused by a positively
charged background density, $\rho_b(x)$, that falls linearly from its bulk value, 
$\overline \rho = \nu_{{\scriptscriptstyle FM}}/2\pi$, 
to zero over a region of width $w$ centered around $x=0$. 
The strength of the confining potential is 
varied by varying the parameter $w$. For small $w$ the 
sharp polarized edge is favored, but as
$w$ increases the edge will reconstruct.

For a spin texture of the form (\ref{eq: edge ansatz}) one finds the following 
gradient, Zeeman and Coulomb energy densities (per unit length of the edge):
\begin{eqnarray}
\label{edgeenergies}
E_{G} &=& \frac{\rho_s}{2} \int_{x_L}^{x_R} dx 
[ k^2(1-f^2)+\frac{f'^2}{1-f^2} ] \ \ \ , 
\nonumber \\
E_{\cal{Z}} &=& 
%-\frac{1}{2}\nu_{{\scriptscriptstyle FM}}\int dx \tilde{g} \rho f=
\frac{\nu_{{\scriptscriptstyle FM}}\tilde{g}}{4 \pi}\left[\frac{k}{4}(1-f_e^2)
-\int_{x_L}^{x_R} dx f\right] \ \ \ ,
\nonumber \\
E_C &=& -\nu_{{\scriptscriptstyle FM}}^2\int_{x_L}^{w/2} \int_{x_L}^{w/2}dx dx'
\delta \rho(x) \delta \rho(x') \ln|x-x'| \ \ \ . \nonumber \\
\end{eqnarray}
Here,  $x_L$ is chosen such that $f(x)=1$ for 
$x\leq x_L$, and
$\delta \rho(x)=\rho(x)-\rho_b(x)$
is the deviation of the electron
density from the background density.

Varying the total energy $E=E_{G}+E_{\cal{Z}} +E_C$ with respect to $f$, we find 
\begin{eqnarray}
\label{edgede}
&f''&(1-f^2) + f f'^2 
\nonumber \\
&+& [k^2 f 
+ \frac{\tilde g \nu_{{\scriptscriptstyle FM}}}{4\pi \rho_s}
+\frac {\nu_{{\scriptscriptstyle FM}}^2 k} {\rho_s}  
\partial V_C/\partial x ] (1-f^2)^2 =0\ \ \ ,
\end{eqnarray}
where
\begin{eqnarray}
V_C(x,f')&=&\int_{x_L}^{x_R} \frac{dx'}{2 \pi} [\rho_b(x')-\overline{\rho}
-\frac{k}{4 \pi} f'(x')] \ln |x-x'|
\nonumber \\
&+&\int_{x_R}^{w/2} \frac{dx'}{2 \pi} \rho_b(x') \ln |x-x'| \ \ \ .
\end{eqnarray} 

We solve Eq. (\ref{edgede}) using the relaxation method described in Sec. 
\ref{sec_relaxation}. 
As an initial ansatz for $f$, we take a modified BP
solution $f_{\scriptscriptstyle BP}(x)=(x^2-4 \lambda^2)/
(x^2+4 \lambda^2)$  ({\it c.f.} Eq. (\ref{sigma solution})) that  obeys the 
boundary 
conditions $f(x_L)=1 , \, f(x_R)=f_e$:
\begin{equation}
\label{edgeansatz}
f_{ans}(x)=f_e + (1-f_e)\frac {1+ f_{\scriptscriptstyle BP}(x_R - x)} 
{1+f_{\scriptscriptstyle BP}(x_R -x_L)} \ \ \ 
.
\end{equation}
The ansatz depends on one parameter $\lambda$ (for given $f_e$).
In the numerical procedure we vary this parameter 
until we get convergence.

We can now study whether 
a spin textured edge has lower energy than the sharp polarized edge for 
given parameters $(\tilde{g},w)$. In particular, we can determine the phase 
transition line in the $(\tilde{g},w)-$plane where the sharp edge reconstructs 
by forming a spin textured edge. Solving  Eq. (\ref{edgede}) using 
the relaxation method for given boundary condition $f_e$ and wave number 
$k$ determines $f(x)$. The energy of this 
state is calculated from Eq. (\ref{edgeenergies}) and $k$ is determined 
by minimizing this energy. This gives the energy of the state 
$E(\tilde g,w,f_e)$. $f_e=1$ gives the energy for the sharp edge. We find that 
the  spin textured edge is lower in energy than the sharp edge for any $f_e$ 
if and only if the derivative of the energy at $f_e=1$ is positive:
$\partial E(\tilde g, w ,f_e)/\partial f_e|_{f_e=1}>0$. The result is that 
the textured edge has lower energy in a triangular
region of parameter space, for small
$\tilde{g}$ and big enough $w$, see \cite{karlhede}.

\section{summary}
\label{sec_sum}
We have used a long wavelength magnetic effective action to compute various
properties of quantum Hall skyrmions and find that these compare favorably
with those obtained by more microscopic Hartree-Fock calculations and by 
analytic small $\tilde g$ expansions. We have
also used the effective action to reliably predict the onset of texturing
at the edges of quantum Hall ferromagnets for small values of the Zeeman
energies. Taken together, these show that this formalism and the relaxational
technique presented here have great utility in the study of quantum Hall
ferromagnets. 

\acknowledgements
We are grateful to Steven Kivelson for useful
discussions and to Daniel Lillieh\"o\"ok for providing Hartree-Fock results. 
This work was supported in part by NSF grant No. DMR-96-32690,
the A. P. Sloan Foundation and the Institute for Advanced Study (SLS),
and the Swedish Natural Science Research Council (AK).


\begin{references}
\bibitem{skothers} There is by now a sizable literature on this subject and
we apologize for not listing much of the recent work except where directly
germane to the topic of this paper.
\bibitem{shoot} The shooting method cannot be used
directly since the origin is a critical point:  The
value of the function and of its first derivative at 
the origin do not specify the solution uniquely.
\bibitem{error} To be precise, we give different weights to 
the errors in the function and in the derivative; this is 
because their typical values differ.
\bibitem{alt} An alternative method would be to use the 
sigma model solution near $r=0$. We would then
have to minimize the energy with respect 
to additional parameters (corresponding to 
unknown derivatives of $f$ at the origin)
and this would necessarily involve several 
relaxation solutions. This is  quite time 
consuming and also has other numerical disadvantages.
\bibitem{enconv} We give the energies to create one skyrmion at fixed 
particle number and fixed magnetic field. These energies are equal for 
the skyrmion and antiskyrmion (with the same value of $|Z|$).
\bibitem{deviation} The deviation is probably partly due to finite size effects in 
the 
Hartree-Fock calculation for which the boundary condition 
is $f(r_{max})=1$ at $r_{max}\approx 32 \ell$ while in the 
effective action calculation we have $r_{max}=600 \ell$.
\end{references}
\end{document}